\documentclass{sig-alternate-05-2015}
\usepackage{graphicx}

\usepackage{epstopdf}

\usepackage{xparse}% http://ctan.org/pkg/xparse
\NewDocumentCommand{\ceil}{s O{} m}{%
  \IfBooleanTF{#1} % starred
    {\left\lceil#3\right\rceil} % \ceil*[..]{..}
    {#2\lceil#3#2\rceil} % \ceil[..]{..}
}

\usepackage{algorithm,algpseudocode}% http://ctan.org/pkg/{algorithms,algorithmicx,caption}

\usepackage[caption=true]{subfig}

\DeclareCaptionType{copyrightbox}

\usepackage{slashbox}

\usepackage[nolist]{acronym}
\begin{acronym}
	\acro{AAA}{Authentication, Authorization and Accounting}
	\acro{ACL}{Access Control List}
	\acro{AKI}{Accountable Key Infrastructure}
	\acro{API}{Application Programming Interface}
    \acro{BSM}{Basic Safety Message}
    \acro{BYOD}{Bring Your Own Device}
	\acro{C2C-CC}{Car2Car Communication Consortium}
	\acro{C2C}{Car-to-Car}
	\acro{C2I}{Car-to-Infrastructure}
	\acro{CA}{Certification Authority}
	\acro{CN}{Common Name}
	\acro{CAM}{Cooperative Awareness Message}
	\acro{CAMP VSC3}{Crash Avoidance Metrics Partnership Vehicle Safety Consortium}
	\acro{CIA}{Confidentiality, Integrity and Availability}
	\acro{CRL}{Certificate Revocation List}
	\acro{CDN}{Content Delivery Network}
	\acro{COCA}{Cornell OnLine Certification Authority}
	\acro{CSR}{Certificate Signing Request}
	\acro{DAA}{Direct Anonymous Attestation}
	\acro{DDoS}{Distributed DoS}
	\acro{DDH}{Decisional Diffie-Helman}
	\acro{DENM}{Decentralized Environmental Notification Message}
	\acro{DHT}{Distributed Hash Table}
	\acro{DL/ECIES}{Discrete Logarithm and Elliptic Curve Integrated Encryption Scheme}
	\acro{DoS}{Denial of Service}
	\acro{DoT}{Department of Transportation}
	\acro{DPA}{Data Protection Agency}
	\acro{DSRC}{Dedicated Short Range Communication}
	\acro{DSS}{Digital Signature Standard}
	\acro{ECU}{Electronic Control Unit}
	\acro{EDR}{Event Data Recorder}
	\acro{ETSI}{European Telecommunications Standards Institute}
	\acro{ECDSA}{Elliptic Curve Digital Signature Algorithm}
	\acro{ECC}{Elliptic Curve Cryptography}
	\acro{EVITA}{E-safety Vehicle Intrusion protected Applications}
	\acro{FOT}{Field Operational Testing}
	\acro{FPGA}{Field-Programmable Gate Array}
	\acro{GPA}{Global Passive Adversary}
	\acro{GN}{GeoNetworking}
	\acro{GS-VLR}{Group Signatures with Verifier Local Revocation}
	\acro{GS}{Group Signatures}
	\acro{GM}{Group Manager}
	\acro{GBA}{Generic Bootstrapping Architecture}
	\acro{GUI}{Graphic User Interface}
	\acro{HSM}{Hardware Security Module}
	\acro{HTTP}{Hypertext Transfer Protocol}
	\acro{IEEE}{Institute of Electrical and Electronics Engineers}
	\acro{IETF}{Internet Engineering Task Force}
	\acro{IoT}{Internet of Things}
	\acro{ITS}{Intelligent Transport System}
	\acro{IT}{Information Technologies}
	\acro{IMSI}{International Mobile Subscriber Identity}
	\acro{IMEI}{International Mobile Station Equipment Identity}
	\acro{IdP}{Identity Provider}
	\acro{IDS}{Intrusion Detection System}
	\acro{ISP}{Internet Service Provider}
	\acro{LEA}{Law Enforcement Agency}
	\acro{LCPP}{Lightweight Conditional Privacy Preservation}
	\acro{LTC}{Long Term Certificate}
	\acro{LTCA}{Long Term \acs{CA}}
	\acro{H-LTCA}{Home-LTCA}
	\acro{LDAP}{Lightweight Directory Access Protocol}
	\acro{LBS}{Location Based Service}
	\acro{LTE}{Long Term Evolution}
	\acro{LuST}{Luxembourg SUMO Traffic}
	\acro{MAC}{Message Authentication Code}
	\acro{MCA}{Message \ac{CA}}
	\acro{MEA}{Misbehavior Evaluation Authority}
	\acro{OBU}{On-Board Unit}
	\acro{OEM}{Original Equipment Manufacturer}
	\acro{OCSP}{Online Certificate Status Protocol}
	\acro{PCA}{Pseudonym \acs{CA}}
	\acro{PDP}{Policy Decision Point}
	\acro{PEP}{Policy Enforcement Point}
	\acro{PIR}{Private Information Retrieval}
	\acro{PKCS}{Public-Key Cryptography Standards}
	\acro{PKI}{Public-Key Infrastructure}
	\acro{PRECIOSA}{Privacy Enabled Capability in Co-operative Systems and Safety Applications}
	\acro{PRESERVE}{Preparing Secure Vehicle-to-X Communication Systems}
	\acro{P2P}{peer-to-peer}
	\acro{PS}{Participatory Sensing}
	\acro{RA}{Resolution Authority}
	\acro{REST}{Representational State Transfer}
	\acro{RBAC}{Role Based Access Control}
	\acro{RCA}{Root Certification Authority}
	\acro{RSU}{Roadside Unit}
	\acro{SAML}{Security Assertion Markup Language}
	\acro{SAS}{Sample Aggregation Service}
	\acro{SCMS}{Security Credential Management System}
	\acro{SCORE@F}{Système COopératif Routier Expérimental Français}
	\acro{SDSI}{Simple Distributed Security Infrastructure}
	\acro{SRAAC}{Secure Revocable Anonymous Authenticated Inter-Vehicle Communication}
	\acro{SeVeCom}{Secure Vehicle Communication}
	\acro{SIT}{Sichere Informationstechnologie}
	\acro{SLC}{Short-Lived Certificate}
	\acro{SoA}{Service-oriented-Approach}
	\acro{SIFS}{Short Inter Frame Space}
	\acro{SSO}{Single-Sign-On}
	\acro{SSL}{Secure Sockets Layer}
	\acro{SOAP}{Simple Object Access Protocol}
	\acro{TACK}{Temporary Anonymous Certified Key}
	\acro{TS}{Task Service}
	\acro{TLS}{Transport Layer Security}
	\acro{TPM}{Trusted Platform Module}
	\acro{TTP}{Trusted Third Party}
	\acro{TVR}{Ticket Validation Repository}
	\acro{URI}{Uniform Resource Identifier}
	\acro{VANET}{Vehicular Ad-hoc Network}
	\acro{V2I}{Vehicle-to-Infrastructure}
	\acro{V2V}{Vehicle-to-Vehicle}
	\acro{V2X}{\ac{V2V} and/or \ac{V2I}}
	\acro{VC}{Vehicular Communication}
	\acro{VM}{Virtual Machine}
	\acro{VSS}{\ac{VC} Security Subsystem}
	\acro{WAVE}{Wireless Access in Vehicular Environments}
	\acro{WSDL}{Web Services Discovery Language}
	\acro{W3C}{World Wide Web Consortium}
	\acro{V}{Vehicle}
	\acro{VANET}{Vehicular Ad-hoc Network}
	\acro{VLR}{Verifier-Local Revocation}
	\acro{VPKI}{Vehicular Public-Key Infrastructure}
	\acro{VM}{Virtual Machine}
	\acro{WS}{Web Service}
	\acro{WoT}{Web of Trust}
	\acro{WSACA}{\ac{WAVE} Service Advertisement \ac{CA}}
	\acro{XML}{Extensible Markup Language}
	\acro{XACML}{eXtensible Access Control Markup Language}
	\acro{3G}{3rd Generation}
\end{acronym}

\usepackage{etoolbox}
\patchcmd{\maketitle}{\@copyrightspace}{}{}{}

\begin{document}
\hyphenation{data-base pseu-do-nyms pseu-do-nym ano-ny-mi-ty a-no-ny-mi-za-tion e-quip-ped pse-udo-ny-mi-ty co-lo-gne in-fra-struc-ture}

\title{Evaluating On-demand Pseudonym Acquisition Policies in \acl{VC} Systems}

\numberofauthors{2} %  in this sample file, there are a *total*
% of EIGHT authors. SIX appear on the 'first-page' (for formatting
% reasons) and the remaining two appear in the \additionalauthors section.
%
\author{
% You can go ahead and credit any number of authors here,
% e.g. one 'row of three' or two rows (consisting of one row of three
% and a second row of one, two or three).
%
% The command \alignauthor (no curly braces needed) should
% precede each author name, affiliation/snail-mail address and
% e-mail address. Additionally, tag each line of
% affiliation/address with \affaddr, and tag the
% e-mail address with \email.
%
% 1st. author
\alignauthor
Mohammad Khodaei\\
       \affaddr{Networked Systems Security Group}\\
       \affaddr{KTH Royal Institute of Technology}\\
       \affaddr{Stockholm, Sweden}\\
       \email{khodaei@kth.se}
% 2nd. author
\and  % use '\and' if you need 'another row' of author names
% 2th. author
Panos Papadimitratos\\
       \affaddr{Networked Systems Security Group}\\
       \affaddr{KTH Royal Institute of Technology}\\
       \affaddr{Stockholm, Sweden}\\
       \email{papadim@kth.se}
}
% There's nothing stopping you putting the seventh, eighth, etc.
% author on the opening page (as the 'third row') but we ask,
% for aesthetic reasons that you place these 'additional authors'
% in the \additional authors block, viz.
%\additionalauthors{Additional authors: John Smith (The Th{\o}rv{\"a}ld Group,
%email: {\texttt{jsmith@affiliation.org}}) and Julius P.~Kumquat
%(The Kumquat Consortium, email: {\texttt{jpkumquat@consortium.net}}).}
%\date{30 July 1999}
% Just remember to make sure that the TOTAL number of authors
% is the number that will appear on the first page PLUS the
% number that will appear in the \additionalauthors section.

\maketitle

\begin{abstract}
Standardization and harmonization efforts have reached a consensus towards using a special-purpose Vehicular Public-Key Infrastructure (VPKI) in upcoming Vehicular Communication (VC) systems. However, there are still several technical challenges with no conclusive answers; one such an important yet open challenge is the acquisition of short-term credentials, \emph{pseudonym}: how should each vehicle interact with the \acs{VPKI}, e.g., how frequently and for how long? Should each vehicle itself determine the pseudonym lifetime? Answering these questions is far from trivial. Each choice can affect both the user privacy and the system performance and possibly, as a result, its security. In this paper, we make a novel systematic effort to address this multifaceted question. We craft three generally applicable policies and experimentally evaluate the \acs{VPKI} system performance, leveraging two large-scale mobility datasets. We consider the most promising, in terms of efficiency, pseudonym acquisition policies; we find that within this class of policies, the most promising policy in terms of privacy protection can be supported with moderate overhead. Moreover, in all cases, this work is the first to provide tangible evidence that the state-of-the-art \acs{VPKI} can serve sizable areas or domain with modest computing resources. 
\end{abstract}


\section{Introduction}
\label{sec:mobihoc-2016-introduction}

\acf{VC} systems aim at enhancing transportation safety and efficiency with a gamut of applications, ranging from collision avoidance to traffic condition updates and \aclp{LBS}~\cite{ETSI-102-638, papadimitratos2009vehicular}. By the same token, the need to enhance security and protect user privacy is well understood \cite{papadimitratos2006securing}. Standardization bodies (IEEE 1609.2 WG \cite{1609draft} and \acs{ETSI}~\cite{ETSI-102-638}) and harmonization efforts (\acs{C2C-CC} \cite{c2c}) have reached a consensus to use public key cryptography: a set of Certification Authorities (CAs) constitute the \acf{VPKI}, providing credentials to legitimate vehicles. Each vehicle, equipped with an \ac{OBU}, is provided with a \ac{LTC} (and has the corresponding private key) to ensure accountable identification of the vehicle. To achieve unlinkability of messages originating a vehicle, a set of short-term certificates, termed \emph{pseudonyms}, are used with the corresponding short-term private keys. A vehicle digitally signs an outgoing message, e.g., a \ac{CAM} or a \ac{DENM}, time- and geo-stamped, using the private key that corresponds to its currently valid pseudonym. It then attaches the pseudonym to the signed messages to facilitate validation by any recipient. Upon reception, the pseudonym is verified (assuming a trust relationship with its issuer) before the message itself (signature validation). This process ensures communication authenticity, integrity and non-repudiation. By frequently changing the pseudonym (and the corresponding private key), the sender privacy is also protected as the pseudonyms per se are inherently unlinkable (if they are issued appropriately, as it will become clear later).

%#######################################
\begin{table*}[!t]
  \centering
  \caption{Comparing Pseudonym Refilling Strategies}
  \label{table:mobihoc-2016-pseudonym-refilling-strategies-comparison}
  \resizebox{18cm}{!} {

  \begin{tabular}{| *{1}c | *{1}{c} | *{1}{c} | *{1}c | *{1}c | } %{lc|c||cc} 
	 	\hline
  		\backslashbox{\textbf{Metrics}}{\textbf{Strategies}}
		& {\textbf{Preloading \& Overlapping}} & {\textbf{Preloading \& Nonoverlapping}} & {\textbf{On-demand \& Overlapping}} & {\textbf{On-demand \& Nonoverlapping}} \\\hline\hline
		Storage size & large & large & small & small \\\hline
		Pseudonym quantity & fixed \& low volume & fixed \& high volume & varying & varying \\\hline
		Pseudonym lifetime & long & short & varying & varying \\\hline
		V-\acs{VPKI} communication frequency & low & low & high & high \\\hline
		Communication overhead & low & low & high & high \\\hline
		Efficient pseudonym utilization & very low & very low & high & high \\\hline
		Pseudonym revocation & difficult \& challenging & difficult \& challenging & no need (lower risk) & no need (lower risk) \\\hline
		Pseudonym vulnerability window & wide & wide & narrow & narrow \\\hline
		Resilience to Sybil-based misbehavior & $\times$ & \checkmark & $\times$ & \checkmark \\\hline
		\shortstack{User privacy protection (probability of linking \\ sets of pseudonyms based on timing information)} & \shortstack{privacy protection: high \\ {(probability of linking: low)}} & \shortstack{privacy protection: low \\ {(probability of linking: high)}} & \shortstack{privacy protection: high \\ {(probability of linking: low)}} & \shortstack{privacy protection: low \\ {(probability of linking: high)}} \\\hline
		\shortstack{User privacy protection (duration for which a pseudonym \\ provider can trivially link sets of pseudonyms for the \\ same vehicle; the longer the duration, the higher \\ the chance to link sets of pseudonyms)} & \shortstack{privacy protection: low \\ {(long duration)} \\ {} \\ {} \\ {}} & \shortstack{privacy protection: low \\ {(long duration)} \\ {} \\ {} \\ {}} & \shortstack{privacy protection: high \\ {(short duration)} \\ {} \\ {} \\ {}} & \shortstack{privacy protection: high \\ {(short duration)} \\ {} \\ {} \\ {}} \\\hline
		Effect on safety application operations & low & low & high & high \\\hline
		Deployment cost (e.g. \acs{RSU}) & low & low & high & high \\\hline
		\shortstack{Proposals \& schemes \\ {}} & \shortstack{{} \\ \acs{C2C-CC}~\cite{c2c}, \acs{PRESERVE}~\cite{preserve-url}, \\ \acs{CAMP VSC3}~\cite{whyte2013security}} & \shortstack{\acs{SeVeCom}~\cite{papadimitratos2008secure}, Safety Pilot~\cite{safety-pilot} \\ {}} & \shortstack{\acs{SRAAC} \cite{fischer2006secure}, V-tokens~\cite{schaub2010v}, \\ CoPRA~\cite{bibmeyer2013copra}} & \shortstack{VeSPA~\cite{vespa2013}, SEROSA~\cite{gisdakis2013serosa}, \\ SR-\acs{VPKI}~\cite{khodaei2014ScalableRobustVPKI}, PUCA~\cite{puca2014}} \\\hline
  \end{tabular}
  }
  \vspace{-0.1em}
\end{table*}
%#######################################

Table \ref{table:mobihoc-2016-pseudonym-refilling-strategies-comparison} classifies different approaches for issuing pseudonyms. A group of proposals suggests preloading the vehicles with the required pseudonyms for a long period, which we term as \emph{preloading} schemes. Accordingly, systems relying on preloading (\acs{C2C-CC}~\cite{c2c}, \acs{CAMP VSC3}~\cite{whyte2013security}, \acs{PRESERVE}~\cite{preserve-url}) issue pseudonyms with \emph{overlapping} lifetimes (validity intervals) to facilitate the operation of the vehicles in safety critical situations. Recall that safety applications necessitate partial linkability to operate because inferring a collision hazard based on unlinkable \acp{CAM} can be hard and error prone. However, having multiple simultaneously valid pseudonyms sets the ground for Sybil-based~\cite{douceur2002sybil} misbehavior, i.e., allowing internal adversaries to inject multiple bogus messages, thus controlling the outcome of specific protocols, e.g., those based on voting~\cite{raya2007eviction}. To thwart this, \acs{SeVeCom}~\cite{papadimitratos2008secure} and Safety Pilot~\cite{safety-pilot} suggested preloading vehicles with pseudonyms that have \emph{non-overlapping} lifetimes (no vehicle has more than one valid pseudonym at any given time). Another group of proposals suggests more frequent vehicles interactions with the \acp{RSU}, i.e., with the \ac{VPKI} servers, e.g., once or multiple times per day, which we term \emph{on-demand} schemes. Among those, \acs{SRAAC}~\cite{fischer2006secure}, V-tokens~\cite{schaub2010v}, and CoPRA~\cite{bibmeyer2013copra} propose issuing pseudonyms with overlapping lifetimes; while VeSPA~\cite{vespa2013}, SEROSA~\cite{gisdakis2013serosa}, SR-\acs{VPKI}~\cite{khodaei2014ScalableRobustVPKI}, and PUCA~\cite{puca2014} propose issuing pseudonyms with non-overlapping lifetimes.

Having different proposals with diverse views on pseudonym acquisition process emphasizes the need to standardize this process with clear objectives. Clearly, there are trade-offs: the longer the interval to obtain pseudonyms, the less frequent the vehicle-\ac{VPKI} communications, but the higher the probability (and the longer the duration) the pseudonym provider can trivially link sets of pseudonyms issued for the same vehicle and thus all communication signed by that vehicle throughout that period. Furthermore, a privacy concern arises for any strategy that issues pseudonyms with non-overlapping lifetimes~\cite{khodaei2014ScalableRobustVPKI}: the use of timing information can enable an eavesdropper to link pseudonyms (the transcript of pseudonymously authenticated messages) by inspecting their successive pseudonym lifetimes. Additionally, efficient pseudonym utilization is a challenging issue: the average trip duration, according to available real mobility traces~\cite{codeca2015lust, uppoor2014generation}, is around 10 minutes during week days. According to the US Census Bureau annual American Community Survey~\cite{acs-survey, whyte2013security}, the average daily commute time is less than an hour. Thus, over-provisioning the vehicles with a large number of unused pseudonyms would be a waste of computation and resources. 

Due to the (i) improved security, i.e., resilience to Sybil-based misbehavior, (ii) user privacy protection, i.e., unlinkable pseudonyms, and (iii) efficiency, i.e., no over-provisioni-ng, on-demand pseudonym acquisition with non-overlapping pseudonym lifetimes is preferable. Within this class of proposals, we seek to address the fundamental questions: \emph{how frequently, and for which period, each vehicle should interact with the \ac{VPKI} to obtain pseudonyms?} Moreover, \emph{should each vehicle have the freedom to determine the pseudonym acquisition periods and the lifetime of the pseudonyms?} We further need \emph{to evaluate the effects of any approach on the overall \ac{VPKI} performance.} The performance of the \ac{VPKI} relates to security and safety: if a pseudonym acquisition approach and specific conditions result in excessive delays to provision a vehicle, then either the vehicle would have to sacrifice its privacy by using its \ac{LTC}, or it would be excluded from the system, exactly reducing safety and transportation efficiency. This is why we need to investigate the overall effect on the \ac{VPKI} system.

\textbf{Contributions:} We make a novel systematic effort to understand the pseudonym acquisition process. We propose three generally applicable policies for pseudonym provisioning that capture alternative approaches in the literature. To evaluate the overall \ac{VPKI} performance, the fundamental metric is the \emph{end-to-end latency} for the vehicle to obtain pseudonyms. This is essentially the only bottleneck any vehicle would encounter in an on-demand approach. We assess the effect of the three policies on the performance of an actual implementation of the most promising, in terms of efficiency, \ac{VPKI}; we leverage two large-scale mobility traces (assuming all vehicles are equipped with \ac{VC} enabling equipments) to emulate realistic conditions. 

In the rest of the paper, we give an overview of the system and pseudonym acquisition policies (Sec. \ref{sec:mobihoc-2016-system-overview}), followed by the description of the protocols (Sec. \ref{sec:mobihoc-2016-pseudonym-refilling-algorithms}). We evaluate the effects of each policy on the overall \ac{VPKI} performance (Sec. \ref{sec:mobihoc-2016-performance-evaluation}) before concluding (Sec. \ref{sec:mobihoc-2016-conclusions}).

\section{System Overview}
\label{sec:mobihoc-2016-system-overview}

%###############################################################################
\begin{table*} [t!]
	\caption{Summary of the Pseudonym Acquisition Policies}
	\centering
	\resizebox{1\textwidth}{!}
		{
		\renewcommand{\arraystretch}{1.15}
			\begin{tabular}{ | c | *{1}{c} | *{1}{c} | c | c | }
				\hline
				\emph{Notation} & \emph{Policy Name} & \emph{Limitations} & \emph{V-\ac{VPKI} Interactions} & \emph{Privacy Preserving} \\\hline
				$P1$  & {User-controlled (user-defined)} & Each user should know the exact trip duration. & once per trip  & {\large $\times$} \\\hline
				\shortstack{$P2$ \\ {} \\ {} \\ {}} & \shortstack{Oblivious \\ {} \\ {} \\ {}} & {\centering \shortstack{Requesting for the last $\Gamma_{P2}$, the user should obtain pseudonyms for the entire duration. \\ {} \\ {} \\ {}}} & \shortstack{\\ {} $\ceil[\bigg]{\dfrac{TripDuration}{\Gamma_{P2}}}$ per trip \\ {}} & \shortstack{\large $\times$ \\ {} \\ {} \\ {}} \\\hline
				\shortstack{$P3$ \\ {} \\ {} \\ {}} & \shortstack{{Universally fixed} \\ {} \\ {}} & {\centering \shortstack{Each user cannot obtain all credentials with a single request. \\ Pseudonym lifetime should be a divisor/factor of the $\Gamma_{P3}$. \\ Requesting for $\Gamma^{i}_{P3}$, each user should obtain pseudonyms for the entire $\Gamma^{i}_{P3}$.}} & \shortstack{$\ceil[\bigg]{\dfrac{TripDuration}{\Gamma_{P3}}}$ per trip \\ {}} & \shortstack{{\large \checkmark} \\ {} \\ {} \\ {}} \\\hline
			\end{tabular}
			\label{table:mobihoc-2016-policy-notation}
		}
   \vspace{-0.75em}
\end{table*}
%###############################################################################

The common denominator among the majority of the proposals and schemes (details in Table \ref{table:mobihoc-2016-pseudonym-refilling-strategies-comparison}) in the literature is essentially a \ac{VPKI} architecture with two main entities: the \ac{LTCA} and the \ac{PCA}. The \ac{LTCA} is responsible for issuing \acp{LTC} for the registered vehicles and is the \emph{policy decision and enforcement point}: it authenticates and authorizes the vehicles. The \ac{PCA} is the responsible authority for issuing the pseudonyms. In case of misbehavior, detected locally~\cite{raya2007eviction} or for other reasons~\cite{Papadi:C:08}, the \ac{RA} initiates a process to resolve, and possibly revoke, a pseudonym based on a set of pseudonymously authenticated messages. Without loss of generality, we follow this common understanding without dwelling on the details of each scheme. We adhere to the assumed adversarial behavior~\cite{papadimitratos2006securing}. We extend it by assuming that the \ac{VPKI} servers are \emph{honest-but-curious}: they correctly execute system protocols and follow the system policies, but they are tempted to harm user privacy. 

%***************************************
\subsection{Pseudonym Acquisition Policies} 
\label{subsec:mobihoc-2016-psnym-acquistion-policies}

The choice of policy for obtaining pseudonyms has diverse ramifications: on the \ac{VPKI} performance and operation as well as the user privacy. The policy determines the volume of the workload (basically, pseudonym requests and related computation and communication latencies) imposed to the \ac{VPKI}. On the other hand, the user privacy is at stake: a transcript of pseudonymously authenticated messages could be linked simply based on the pseudonym lifetime and issuance times \cite{khodaei2014ScalableRobustVPKI}, and requests could act as user \emph{``fingerprints''}. Simply put, individually determined pseudonym lifetimes allow an observer to link pseudonyms of the same vehicle only based on timing information of the credentials (without even examining the content of the message). To systematically investigate the effect of diverse on-demand non-overlapping pseudonym acquisition methods, we define three representatives, summarized in Table \ref{table:mobihoc-2016-policy-notation}.

\vspace{1em}
\textbf{User-controlled (user-defined) policy (P1):} A vehicle requests pseudonyms for its residual (ideally entire) trip duration at the start of trip. We presume each vehicle \emph{precisely estimates} the trip duration in advance, e.g., based on automotive navigation systems, previous trips, or user input. The \ac{PCA} determines the pseudonym lifetime, either fixed for all vehicles or flexible for each requester. Additional pseudonyms should be requested if the actual trip duration exceeds the estimated one to ensure that the vehicle is always equipped with enough valid pseudonyms.

\textbf{Oblivious policy (P2):} The vehicle interacts with the \ac{VPKI} every $\Gamma_{P2}$ seconds (determined by the \ac{PCA} and fixed for all users) and it requests pseudonyms for the entire $\Gamma_{P2}$ time interval until the vehicle reaches its destination. This results in over-provisioning of pseudonyms only during the last iteration.\footnote{As an optimization, a vehicle could \emph{roughly estimate} the residual trip duration, $D_{v}$; if $D_{v}\ll\Gamma_{P2}$, then it will request only for the $D_{v}$ period rather than the entire $\Gamma_{P2}$ interval; otherwise, it will request for the $\Gamma_{P2}$ interval. Here for simplicity, we do not consider this optimization.} The difference, in comparison to P1, is that either the vehicle does not know the exact trip duration, or, it does not attempt to estimate, or possibly, overestimate it; thus, P2 is oblivious to the trip duration. 

\textbf{Universally fixed policy (P3):} The \ac{PCA} has predetermined \emph{``universally''} fixed interval, $\Gamma_{P3}$, and pseudonym lifetime, $\tau_{P}$. At the start of its trip, a vehicle requests pseudonyms for the \emph{``current''} $\Gamma_{P3}$, out of which useful (non-expired) ones are actually obtained for the residual trip duration. For the remainder of the trip, the vehicle requests pseudonyms for the entire $\Gamma_{P3}$ at each time. This policy issues time-aligned pseudonyms for all vehicles; thus, timing information does not harm user privacy.

\begin{figure} [!t] %[t!] %[!htbp]%[h!]%[!htbp]%[h!]%[htp]
	\centering
   	\includegraphics[trim=8cm 0cm 0cm 7cm, clip=true, totalheight=0.45\textheight, width=0.45\textwidth, angle=0, keepaspectratio] {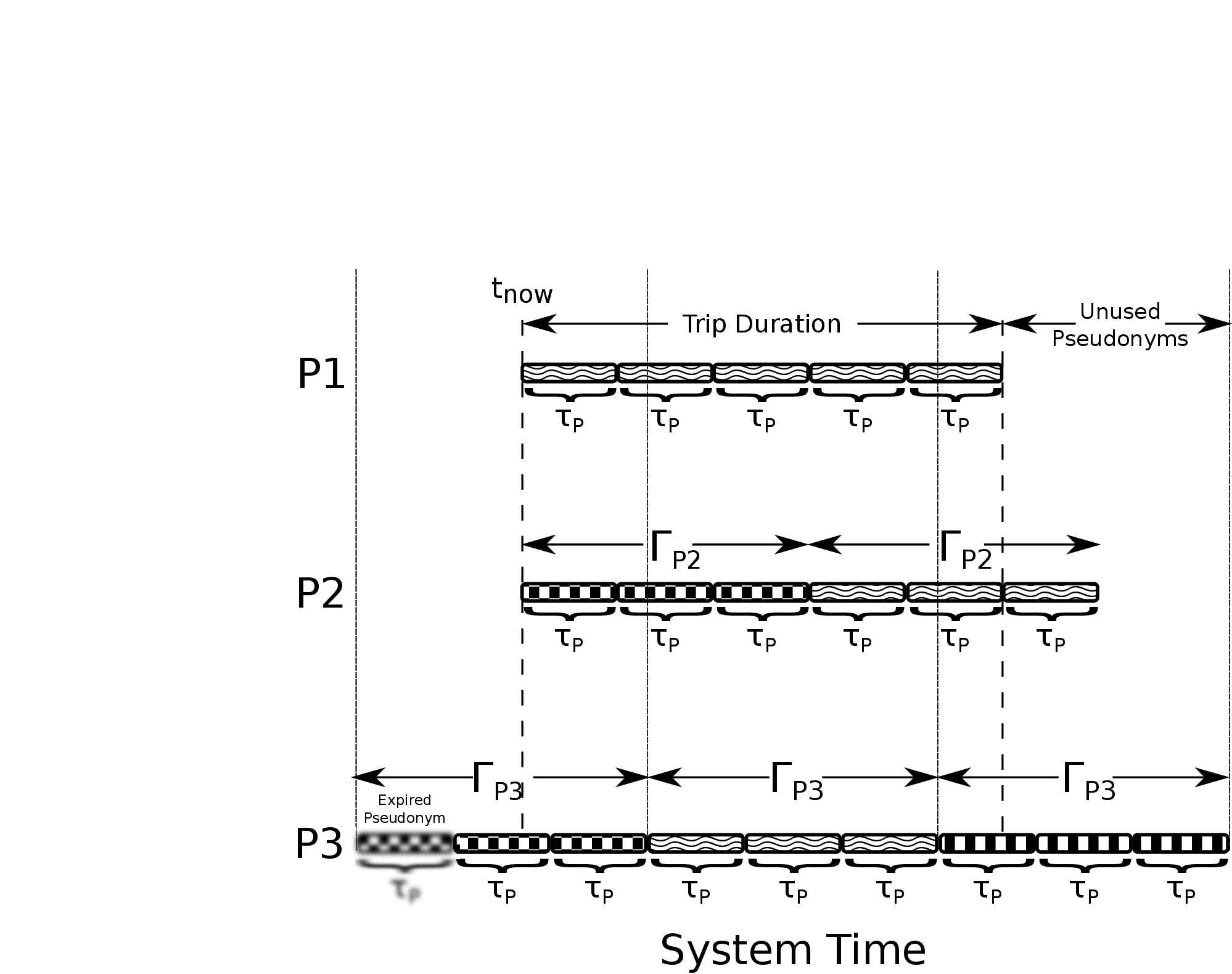}
   \caption{A Schematic Comparison of P1,$\:$P2,$\:$and$\:$P3} %and 
   \label{fig:mobihoc-2016-pseudonym-acquistion-policies-schematic}
   \vspace{-0.5em}
\end{figure}

Fig. \ref{fig:mobihoc-2016-pseudonym-acquistion-policies-schematic} illustrates the three pseudonym acquisition policies with respect to trip duration, pseudonym lifetime ($\tau_{P}$), and the \ac{PCA}-determined periods ($\Gamma_{P2}$ and $\Gamma_{P3}$). Using P1 and P2, the exact time of requests and all subsequent requests until the end of trip could be unique, or one of few, and thus linkable even by an external observer; it might be unlikely in a specific region to have multiple requests at a specific instance. While in contrast for P3, the requesting intervals fall within the \emph{``universally''} fixed interval ($\Gamma_{P3}$) and the issued pseudonyms are aligned with the global system time (\ac{PCA} clock); therefore, at any point in time all vehicles in a given area will be transmitting under pseudonyms which are indistinguishable, one from another, based on timing information alone; thus, protecting user privacy.

%#######################################
%#######################################

\section{Pseudonym Acquisition}
\label{sec:mobihoc-2016-pseudonym-refilling-algorithms}

%######################################
\begin{table} [t!] %[!h]
	\caption{Notation used in the protocols}
	\centering
	\resizebox{0.465\textwidth}{!}
		{
		\renewcommand{\arraystretch}{1.1}
		\begin{tabular}{l | *{1}{c} r}
			\hline \hline
			$P$, $P^{i}_{v}$, $(P^{i}_{v})_{\sigma_{PCA}}$  & \emph{current valid pseudonym signed by the \acs{PCA}} \\\hline % or $(P)_{PCA}$ 
			$(LK_v, Lk_v)$ & \emph{long-term public and private key pairs} \\\hline
			$(K^i_v, k^i_v)$ & \emph{current valid pseudonymous public and private key pairs} \\\hline
			$Id_{req}, Id_{res}$ & \emph{request/response identifiers} \\ \hline
			$Id_{CA}$  & \emph{\acl{CA} unique identifier} \\\hline
			$(msg)_{\sigma_{v}}$ & \emph{a signed message with the vehicle's private key} \\\hline
			$N$ & \emph{a nonce} \\\hline
			$t_{now}, t_s, t_e, t_{date}$ & \emph{current, starting, ending, and a specific day timestamps} \\\hline %fresh/
			$SN$ & \emph{serial number} \\\hline
			$Exp_{tkt}$ & \emph{ticket expiration time} \\\hline
			$H()$ & \emph{hash function} \\\hline
			$Sign(Lk_{ca}, msg)$ & \emph{signing a message with private key ($Lk$) of the \acs{CA}} \\\hline %process of 
			$Verify(LTC_{ca}, msg)$ & \emph{verifying a message with the \acs{CA}'s public key} \\ \hline %process of 
			$\tau_{P}$ & \emph{pseudonym lifetime} \\\hline
			$\Gamma_{Px}$ & \emph{interacting period/interval with the \ac{VPKI} for policy x} \\\hline % duration interaction 
			$IK$ & \emph{identifiable key} \\\hline
			$\zeta$ & \emph{a temporary variable} \\\hline
			\hline
		\end{tabular}
		\renewcommand{\arraystretch}{1}
		\label{table:mobihoc-2016-protocols-notation}
		}
	\vspace{-1em}
\end{table}
%######################################

For completeness and broader applicability, we extend and refine the common system model (Sec. \ref{sec:mobihoc-2016-system-overview}) to enable multi-domain \ac{VPKI} operation \cite{khodaei2014ScalableRobustVPKI}. Our system works as follows: each registered vehicle interacts first with its \ac{LTCA} to obtain a ticket. The \ac{LTCA} authenticates and authorizes the requester, issuing a \emph{service-granting ticket} for the requester, which enables the vehicle to obtain pseudonyms from any \ac{PCA}. The trust establishment between the \ac{LTCA} and the \ac{PCA} is with the help of a higher-level authority or by using cross certification \cite{khodaei2015VTMagazine}. The vehicle, i.e., the \ac{OBU}\footnote{The terms vehicle and \ac{OBU} are used interchangeably.}, decides when to trigger the pseudonym acquisition process based on different parameters, e.g., the number of remaining valid pseudonyms, the residual trip duration, and the networking connectivity. We presume connectivity to the \ac{VPKI} (via \acp{RSU}, cellular networks, or other connectivities) to execute all of these protocols. Should the connectivity be intermittent, the \ac{OBU} could initiate the protocols proactively when there is connectivity. Further discussion on a reliable connectivity to the \ac{VPKI} is beyond the scope of this paper. The notation used in the protocols is given in Table \ref{table:mobihoc-2016-protocols-notation}.

\textbf{Ticket Acquisition Process (Protocols \ref{protocol:mobihoc-2016-requesting-ticket-algorithm} and \ref{protocol:mobihoc-2016-issuing-ticket-algorithm}):} Assume the \ac{OBU} decides to obtain pseudonyms from a specific \ac{PCA}. If the relevant policy is P1, each vehicle \emph{estimates} the actual trip duration $[t_{s}, \: t_{e}]$ (steps \ref{protocol:mobihoc-2016-requesting-ticket-algorithm}.2\textendash\ref{protocol:mobihoc-2016-requesting-ticket-algorithm}.3, i.e., steps 2\textendash3 in protocol \ref{protocol:mobihoc-2016-requesting-ticket-algorithm}) while with P2, each vehicle requests pseudonyms for $[t_{s}, \: t_{s}+\Gamma_{P2}]$ (steps \ref{protocol:mobihoc-2016-requesting-ticket-algorithm}.4\textendash\ref{protocol:mobihoc-2016-requesting-ticket-algorithm}.5). If the relevant policy is P3, the vehicle calculates the trip duration based on the date of travel, $t_{date}$, and the exact time of travel corresponding to the universally fixed interval, $\Gamma_{P3}$, of that specific \ac{PCA} (steps \ref{protocol:mobihoc-2016-requesting-ticket-algorithm}.6\textendash\ref{protocol:mobihoc-2016-requesting-ticket-algorithm}.7). It then calculates the hash value of the concatenation of the specific \ac{PCA} identity with a random number; this conceals the identity of the \ac{PCA} from the \ac{LTCA}. The vehicle prepares the request (step \ref{protocol:mobihoc-2016-requesting-ticket-algorithm}.9) before signing it using the private key corresponding to its \ac{LTC} (step \ref{protocol:mobihoc-2016-requesting-ticket-algorithm}.10), and returning the ticket request (step \ref{protocol:mobihoc-2016-requesting-ticket-algorithm}.11). It will then interact with the \ac{LTCA} over a bidirectional authenticated \ac{TLS}. 

%######################################
\begin{algorithm}[t!]
	\floatname{algorithm}{Protocol}
	\caption{Ticket Request (from the \acs{LTCA})}
	\label{protocol:mobihoc-2016-requesting-ticket-algorithm}
	\begin{algorithmic}[1]
		\Procedure{ReqTicket}{$P_{x}, \Gamma_{Px}, t_{s}, t_{e}, t_{date}$}
			\If {$P_{x} = P1$}
				\State $(t_s, t_e) \gets (t_{s}, t_{e})$
			\ElsIf {$P_{x} = P2$}
				%\State $(t_s, t_e) \gets (D_v\ll\Gamma_{P2}) ? (t_s, D_v) : (t_s, \Gamma_{P2})$
				\State $(t_s, t_e) \gets (t_s, t_{s}+\Gamma_{P2})$
			\ElsIf {$P_{x} = P3$}
				\State $(t_s, t_e) \gets (t_{date}+\Gamma^{i}_{P3}), t_{date}+\Gamma^{i+1}_{P3})$
			\EndIf
			\State $\zeta \leftarrow (Id_{tkt\textnormal{-}req}, H(Id_{PCA}\|Rnd_{tkt}), t_s, t_e)$
			\State $(\zeta)_{\sigma_{v}} \leftarrow Sign(Lk_v, \zeta)$ 
			\State \textbf{return} $((\zeta)_{\sigma_{v}}, \acs{LTC}_v, N, t_{now})$
		\EndProcedure
	\end{algorithmic}
\end{algorithm}
%######################################

Upon reception of the ticket request, the \ac{LTCA} verifies the \ac{LTC} (thus authenticating and authorizing the requester) and the signed message (step \ref{protocol:mobihoc-2016-issuing-ticket-algorithm}.2). The \ac{LTCA} calculates the \emph{``ticket identifiable key''} to bind the ticket to the \ac{LTC} as: $H(\acs{LTC}_v || t_s || t_e || Rnd_{IK_{tkt}})$ (step \ref{protocol:mobihoc-2016-issuing-ticket-algorithm}.3); this prevents a compromised \ac{LTCA} from mapping a different \ac{LTC} during the resolution process. The \ac{LTCA} then encapsulates (step \ref{protocol:mobihoc-2016-issuing-ticket-algorithm}.4), signs (step \ref{protocol:mobihoc-2016-issuing-ticket-algorithm}.5), and delivers the response (step \ref{protocol:mobihoc-2016-issuing-ticket-algorithm}.6). 

%######################################
\begin{algorithm}[t!]
	\floatname{algorithm}{Protocol}
	\caption{Issuing a Ticket (by the \acs{LTCA})}
	\label{protocol:mobihoc-2016-issuing-ticket-algorithm}
	\begin{algorithmic}[1]
		\Procedure{IssueTicket}{$(msg)_{\sigma_{v}}, \acs{LTC}_v, N, t_{now}$}
			\State $\text{Verify}(\acs{LTC}_v, (msg)_{\sigma_{v}})$
			\State ${IK_{tkt} \leftarrow H(\acs{LTC}_v || t_s || t_e || Rnd_{IK_{tkt}})}$ 
			\State $\zeta \leftarrow (SN,H(Id_{PCA}\|Rnd_{tkt}),IK_{tkt},t_s,t_e,Exp_{tkt})$
			%\Statex $t_s,t_e,Exp_{tkt})$
			\State $(tkt)_{\sigma_{ltca}} \leftarrow Sign(Lk_{ltca}, \zeta)$
			\State \textbf{return} $((tkt)_{\sigma_{ltca}}, Rnd_{IK_{tkt}}, N+1, t_{now})$
		\EndProcedure
	\end{algorithmic}
\end{algorithm}

%######################################

\textbf{Pseudonym Acquisition Process (Protocols \ref{protocol:mobihoc-2016-requesting-psnyms-algorithm} and \ref{protocol:mobihoc-2016-issuing-psnyms-algorithm}):} Upon reception of a valid ticket, the vehicle generates required \ac{ECDSA} public/private key pairs (steps \ref{protocol:mobihoc-2016-requesting-psnyms-algorithm}.2\textendash\ref{protocol:mobihoc-2016-requesting-psnyms-algorithm}.6). It then prepares the request (step \ref{protocol:mobihoc-2016-requesting-psnyms-algorithm}.7) and sends the pseudonym request to the \ac{PCA} over a unidirectional authenticated \ac{TLS}. 

Having received a request, the \ac{PCA} verifies the ticket, signed by the \ac{LTCA} (assuming a trust is established) (steps \ref{protocol:mobihoc-2016-issuing-psnyms-algorithm}.2\textendash\ref{protocol:mobihoc-2016-issuing-psnyms-algorithm}.3). The \ac{PCA} then verifies the pseudonym provider identity (step \ref{protocol:mobihoc-2016-issuing-psnyms-algorithm}.4) and the requesting intervals for obtaining pseudonyms (step \ref{protocol:mobihoc-2016-issuing-psnyms-algorithm}.5). Afterward, the \ac{PCA} initiates a proof-of-possession protocol to verify the ownership of the corresponding private keys. Then, it calculates the \emph{``pseudonym identifiable key''} to bind the pseudonyms to the ticket as: $H(IK_{tkt} || K^i_v || t_{s}^i || t_{e}^i || Rnd_{IK^{i}_{v}})$. This essentially prevents a compromised \ac{PCA} from mapping a different ticket during resolution process. It issues the pseudonyms (steps \ref{protocol:mobihoc-2016-issuing-psnyms-algorithm}.6\textendash\ref{protocol:mobihoc-2016-issuing-psnyms-algorithm}.12) and delivers the response (step \ref{protocol:mobihoc-2016-issuing-psnyms-algorithm}.13).

%######################################
\begin{algorithm} [t!]
	\floatname{algorithm}{Protocol}
	\caption{Pseudonym Request (from the \acs{PCA})}
	\label{protocol:mobihoc-2016-requesting-psnyms-algorithm}
	\algloop{For}{}
	\algblock{Begin}{End}
	\begin{algorithmic}[1]
		\Procedure{ReqPsnyms}{$t_{s}, t_{e}, (tkt)_{\sigma_{ltca}}$} %\Gamma_{Px}, P_{x}, 
			\For{i:=1 to \textbf{n}}{}
				\Begin
					\State $\text{Generate}(K^{i}_{v}, k^i_v)$  
					\State $(K^i_v)_{\sigma_{k^i_v}} \gets \text{Sign}(k^i_v, K^{i}_{v})$ % Proof-of-possession 
				\End
			\State $psnymReq \gets {(Id_{req}, Rnd_{tkt}, t_{s}, t_{e}, (tkt)_{\sigma_{ltca}},}$
			\Statex $\hspace{7.8em}{\{(K^1_v)_{\sigma_{k^1_v}}, ..., (K^n_v)_{\sigma_{k^n_v}}\}, N, t_{now})}$
			\State \textbf{return} $psnymReq$
		\EndProcedure
	\end{algorithmic}
\end{algorithm}
%######################################

%######################################
\begin{algorithm}[t!]
	\floatname{algorithm}{Protocol}
	\caption{Issuing Pseudonyms (by the \acs{PCA})}
	\label{protocol:mobihoc-2016-issuing-psnyms-algorithm}
	\algloop{For}{}
	\algblock{Begin}{End}
	\begin{algorithmic}[1]
		\Procedure{IssuePsnyms}{$psnymReq$}
			\State $psnymReq \to {(Id_{req}, Rnd_{tkt}, t_{s}, t_{e}, (tkt)_{\sigma_{ltca}},}$
			\Statex $\hspace{8em} {\{(K^1_v)_{\sigma_{k^1_v}}, ..., (K^n_v)_{\sigma_{k^n_v}}\}, N, t_{now})}$
			\State $\text{Verify}(LTC_{ltca}, (tkt)_{\sigma_{ltca}})$ 
			\State $H(Id_{this\textnormal{-}PCA}\|Rnd_{tkt}) \stackrel{?}{=} H(Id_{PCA}\|Rnd_{tkt})$ 
			\State ${[t_{s}, \: t_{e}] \overset{?}{=} ([t_s, t_e])_{tkt}}$ 
			\For{i:=1 to \textbf{n}}{}
				\Begin
					\State $\text{Verify}(K^{i}_{v}, (K^i_v)_{\sigma_{k^i_v}})$ % Proof-of-possession \notag 
					\State ${IK_{P^i} \gets H(IK_{tkt} || K^i_v || t_{s}^i || t_{e}^i|| Rnd_{IK^{i}_{v}}})$
					\State ${\zeta \leftarrow (SN^i, K^i_v, IK_{P^i}, t_{s}^i, t_{e}^i)}$
					\State $(P^i_v)_{\sigma_{pca}} \leftarrow Sign(Lk_{pca}, \zeta)$
				\End
			%\State \textbf{return} $(\{(P^1_v)_{\sigma_{pca}}, \dots, (P^n_v)_{\sigma_{pca}}\}, N\textnormal{+}1, t_{now}) $
			\State \textbf{return} $(\{(P^1_v)_{\sigma_{pca}}, \dots, (P^n_v)_{\sigma_{pca}}\}, $
			\Statex $\hspace{5.5em}\{Rnd_{IK^{1}_{v}}, \dots, Rnd_{IK^{n}_{v}}\},  N\textnormal{+}1, t_{now}) $
		\EndProcedure
	\end{algorithmic}
\end{algorithm}

%######################################

\section{Performance Evaluation}
\label{sec:mobihoc-2016-performance-evaluation}

Due to the lack of a large-scale deployment of \ac{VC} systems, we resort to realistic large-scale mobility traces, which determine the period the vehicles need pseudonyms. We extract two features of interest from the mobility traces, i.e., departure time and trip duration, and we apply policies described in Sec. \ref{subsec:mobihoc-2016-psnym-acquistion-policies} to assess the efficiency of the full-blown implementation of our \ac{VPKI} for a large-scale deployment. The main metric is the \emph{end-to-end pseudonym acquisition latency}, i.e., the delay from the initialization of protocol \ref{protocol:mobihoc-2016-requesting-ticket-algorithm} till the successful completion of protocol \ref{protocol:mobihoc-2016-issuing-psnyms-algorithm} (steps \ref{protocol:mobihoc-2016-requesting-ticket-algorithm}.1\textendash\ref{protocol:mobihoc-2016-issuing-psnyms-algorithm}.14), measured at the vehicle. The processing time to generate the key pairs (steps \ref{protocol:mobihoc-2016-requesting-psnyms-algorithm}.2\textendash\ref{protocol:mobihoc-2016-requesting-psnyms-algorithm}.6) is not considered here as the \ac{OBU} can generate them off-line. 

\subsection{Experimental Setup}
\label{subsec:mobihoc-2016-experimental-setup}

\textbf{\ac{VPKI} Testbed:} We create a testbed comprising different \acp{VM} allocated to distinct \ac{VPKI} servers; Table \ref{table:mobihoc-2016-vm-SIS-clients-specifications} details the specifications for the servers and the emulated client\footnote{The processing power of the client is comparable to the Nexcom boxes (dual-core 1.66GHz, 2GB memory) in \acs{PRESERVE} project~\cite{preserve-url} as we execute all clients in one \ac{VM}.}. Our implementation is in C++ and we use OpenSSL for the cryptographic protocols and primitives, i.e., \ac{TLS} and \ac{ECDSA}-256 (according to the standards~\cite{ETSI-102-638, 1609draft}). We run the experiments in our testbed with \ac{VPKI} servers and clients (emulating \acp{OBU}) running on the \acp{VM}. This set up eliminates the network propagation delays of \ac{OBU}-\ac{VPKI} connectivity. Depending on the actual \ac{OBU}-\ac{VPKI} connectivity, the network propagation delays would vary; for simplicity, we do not consider it here. 

\textbf{Mobility Traces:} To have realistic arriving requests to the \ac{VPKI}, we used two microscopic mobility vehicle datasets: TAPASCologne \cite{uppoor2014generation} and \ac{LuST} \cite{codeca2015lust}, detailed in Table \ref{table:mobihoc-2016-mobility-traces-information}. The former one represents the traffic demand information across the K\"oln urban area (available for 2 hours, 6-8 AM) while the latter presents a full-day realistic mobility pattern in the city of Luxembourg. 

%#######################################
\begin{table}[!t] %hb
	\centering
		\caption{Servers and Clients Specifications} 
		\resizebox{0.45\textwidth}{!}
		{
			\begin{tabular}{l | *{4}{c} r}
				& \ac{LTCA} & \ac{PCA} & \ac{RA} & Client \\ \hline %& NexCom [TBD!] 
				Number of entities & 1 & 1 & 1 & 1 \\ % & 25
				Dual-core CPU (Ghz) & 2.0 & 2.0 & 2.0 & 2.0 \\ %& 1.66 
				BogoMips          & 4000 & 4000 & 4000 & 4000 \\ %& 3333.36
				Memory            & 2GB & 2GB & 1GB & 1GB \\ %& 1GB
				Database          & MySQL & MySQL & MySQL & MySQL \\ %& MySQL 
			\end{tabular}
			\label{table:mobihoc-2016-vm-SIS-clients-specifications}
		}
\end{table}
%#######################################

%#######################################
\begin{table}[!t] %hb
	\centering
	\caption{Mobility Traces Information}
	\resizebox{0.45\textwidth}{!}
    {
		\begin{tabular}{l | *{4}{c} r}
			& TAPASCologne & \acs{LuST} \\ \hline
			Number of vehicles & 75,576 & 138,259 \\
			Number of trips  & 75,576 & 287,939 \\ 
			Duration of snapshot (hour) & 24 & 24 \\
			Available duration of snapshot (hour) & 2 (6-8 AM) & 24 \\
			Average trip duration (sec.) & 590.49 & 692.81 \\ % 590.498663597 & 692.816627827
		\end{tabular}
		\label{table:mobihoc-2016-mobility-traces-information}
	}
	\vspace{-0.5em}
\end{table}
%#######################################

\subsection{\ac{VPKI} Servers Performance}

\begin{figure} [!t] %[t!] %[!htbp]%[h!]%[!htbp]%[h!]%[htp]
   \vspace{-0.5em}
	\centering
   \subfloat[TAPASCologne dataset]{ 
		\hspace{-2em}\includegraphics[trim=0cm 0cm 0cm 0cm, clip=true, totalheight=0.29\textheight, width=0.29\textwidth, angle=0, keepaspectratio] {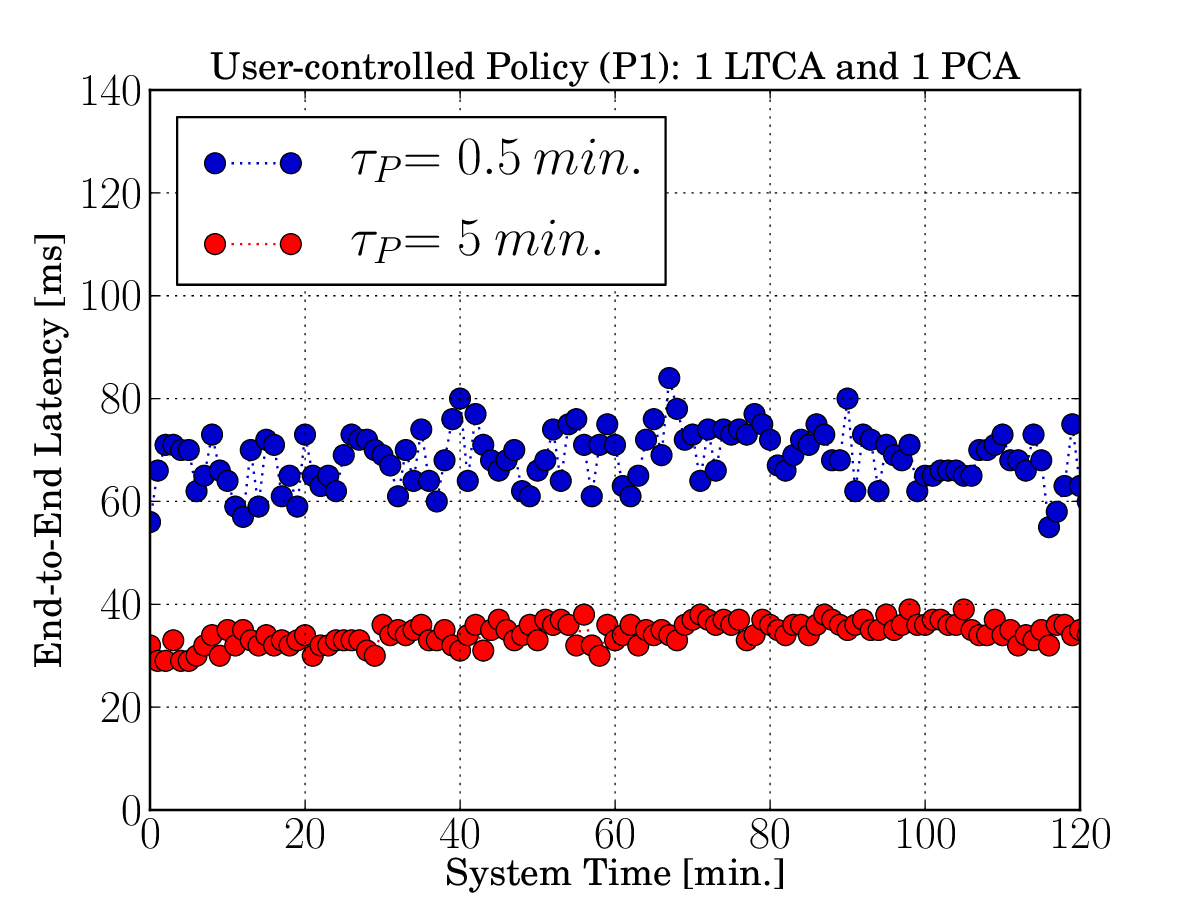}}
	\subfloat[\ac{LuST} dataset]{ 
		\hspace{-1.5em} \includegraphics[trim=0cm 0cm 0cm 0cm, clip=true, totalheight=0.29\textheight, width=0.29\textwidth, angle=0, keepaspectratio] {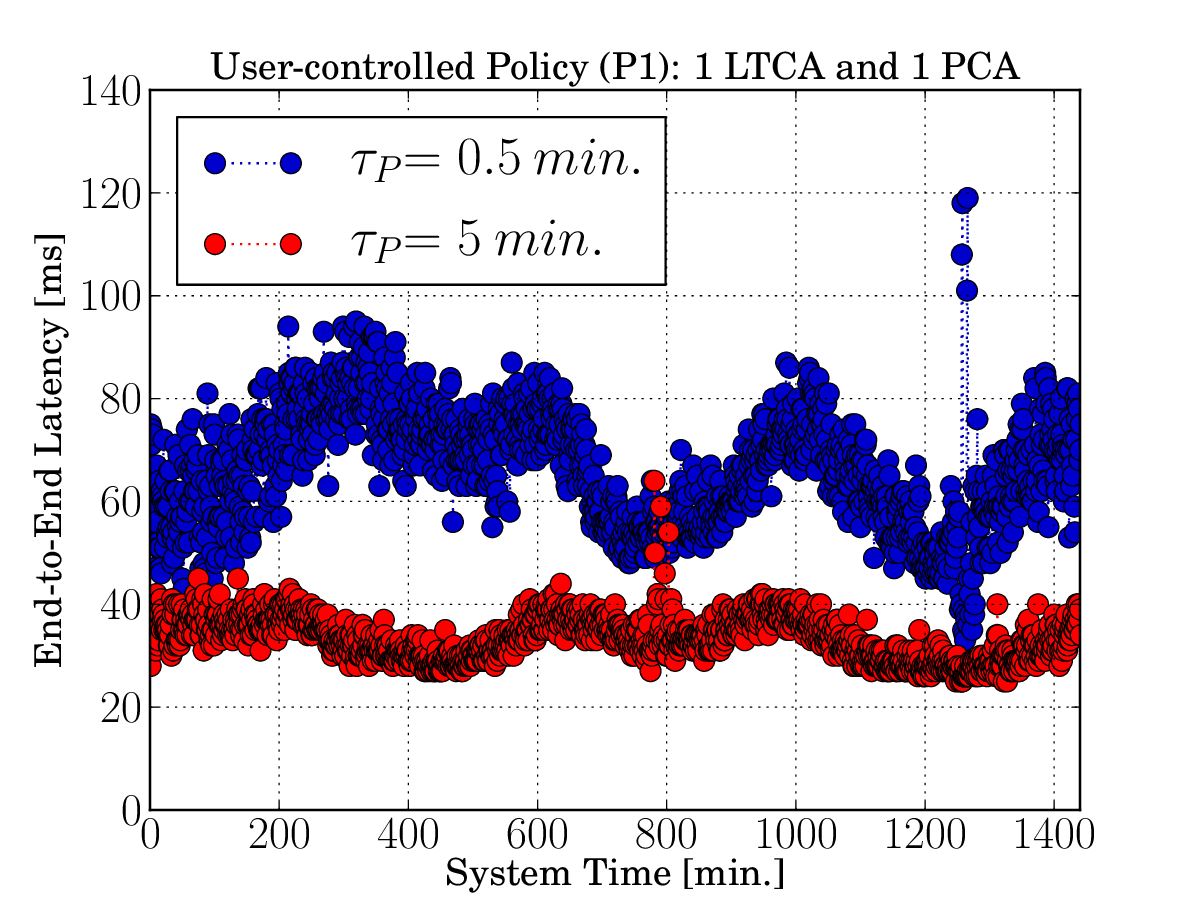}}
   \caption{End-to-end latency for P1}
   \label{fig:mobihoc-2016-tapas_lust_entire_vehicle_delay_over_time_usercontrolled_policy}
   \vspace{-1em}
\end{figure}

\begin{figure} [!t] %[t!] %[!htbp]%[h!]%[!htbp]%[h!]%[htp]
   \vspace{-0.5em}
	\centering
   \subfloat[TAPASCologne dataset]{ 
		\hspace{-2em}\includegraphics[trim=0cm 0cm 0cm 0cm, clip=true, totalheight=0.29\textheight, width=0.29\textwidth, angle=0, keepaspectratio] {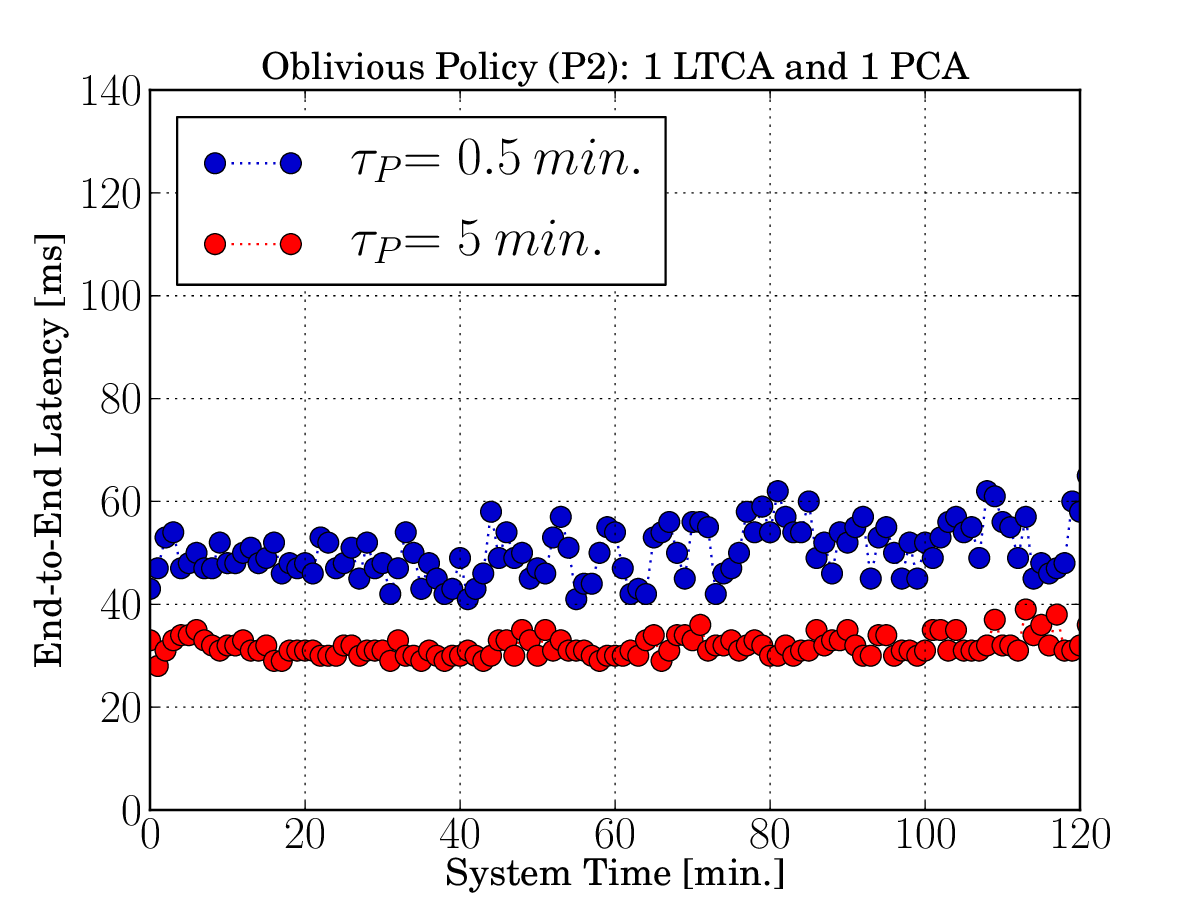}}
	\subfloat[\ac{LuST} dataset]{ 
		\hspace{-1.5em}\includegraphics[trim=0cm 0cm 0cm 0cm, clip=true, totalheight=0.29\textheight, width=0.29\textwidth, angle=0, keepaspectratio] {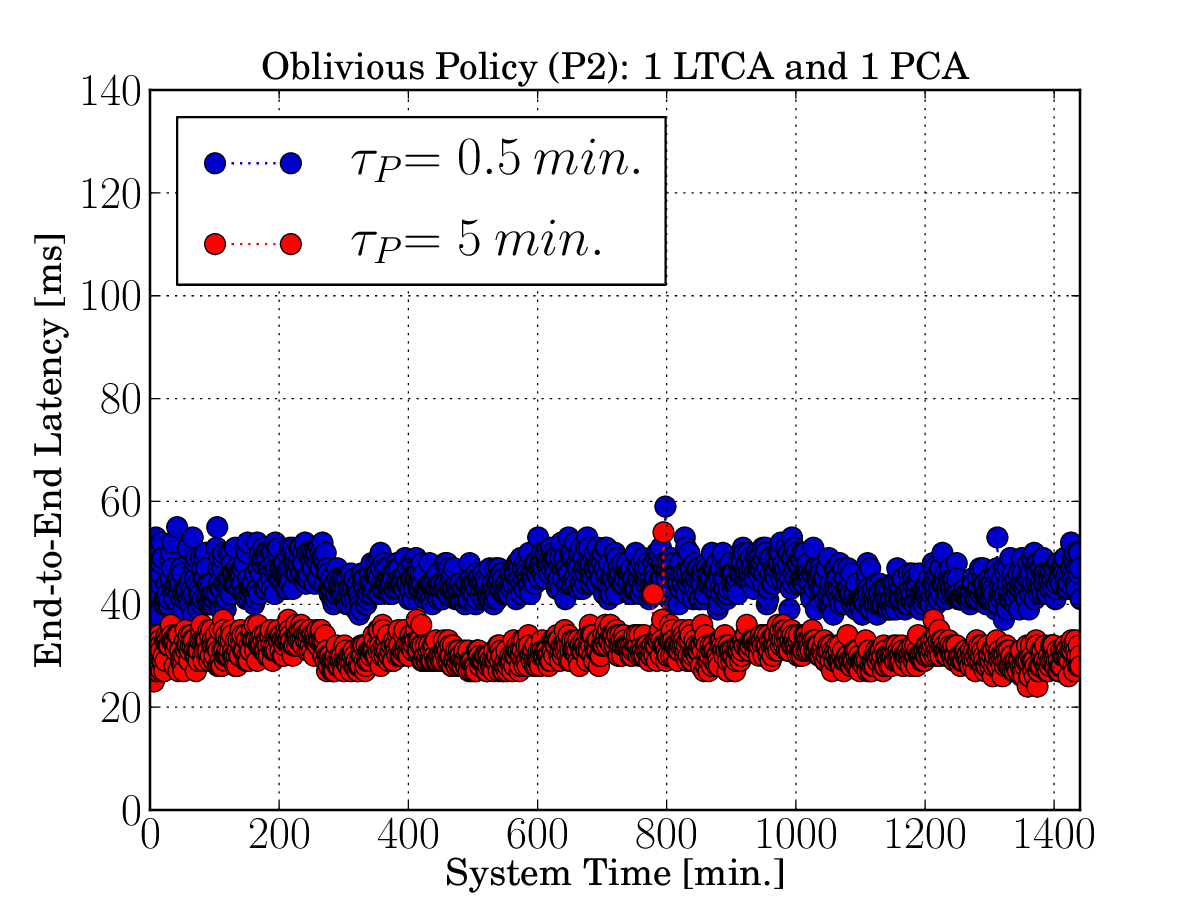}}
   \caption{End-to-end latency for P2, $\Gamma_{P2}$=5 minutes}
   \label{fig:mobihoc-2016-tapas_entire_vehicle_delay_over_time_upperlimit_policy}
   \vspace{-1em}
\end{figure}

\begin{figure} [!t] %[t!] %[!htbp]%[h!]%[!htbp]%[h!]%[htp]
   \vspace{-0.5em}
	\centering
   \subfloat[TAPASCologne dataset]{ 
		\hspace{-2em}\includegraphics[trim=0cm 0cm 0cm 0cm, clip=true, totalheight=0.29\textheight, width=0.29\textwidth, angle=0, keepaspectratio] {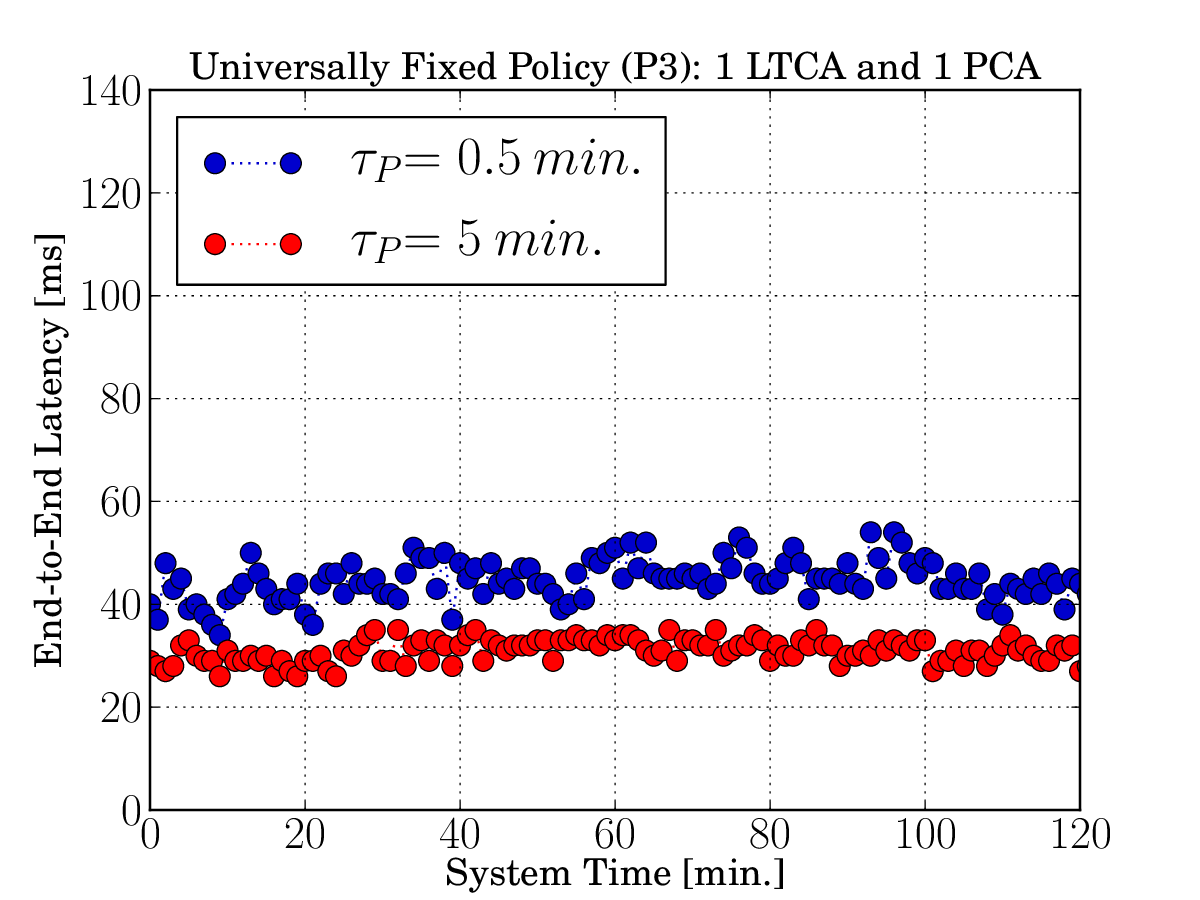}}
	\subfloat[\ac{LuST} dataset]{ 
		\hspace{-1.5em}\includegraphics[trim=0cm 0cm 0cm 0cm, clip=true, totalheight=0.29\textheight, width=0.29\textwidth, angle=0, keepaspectratio] {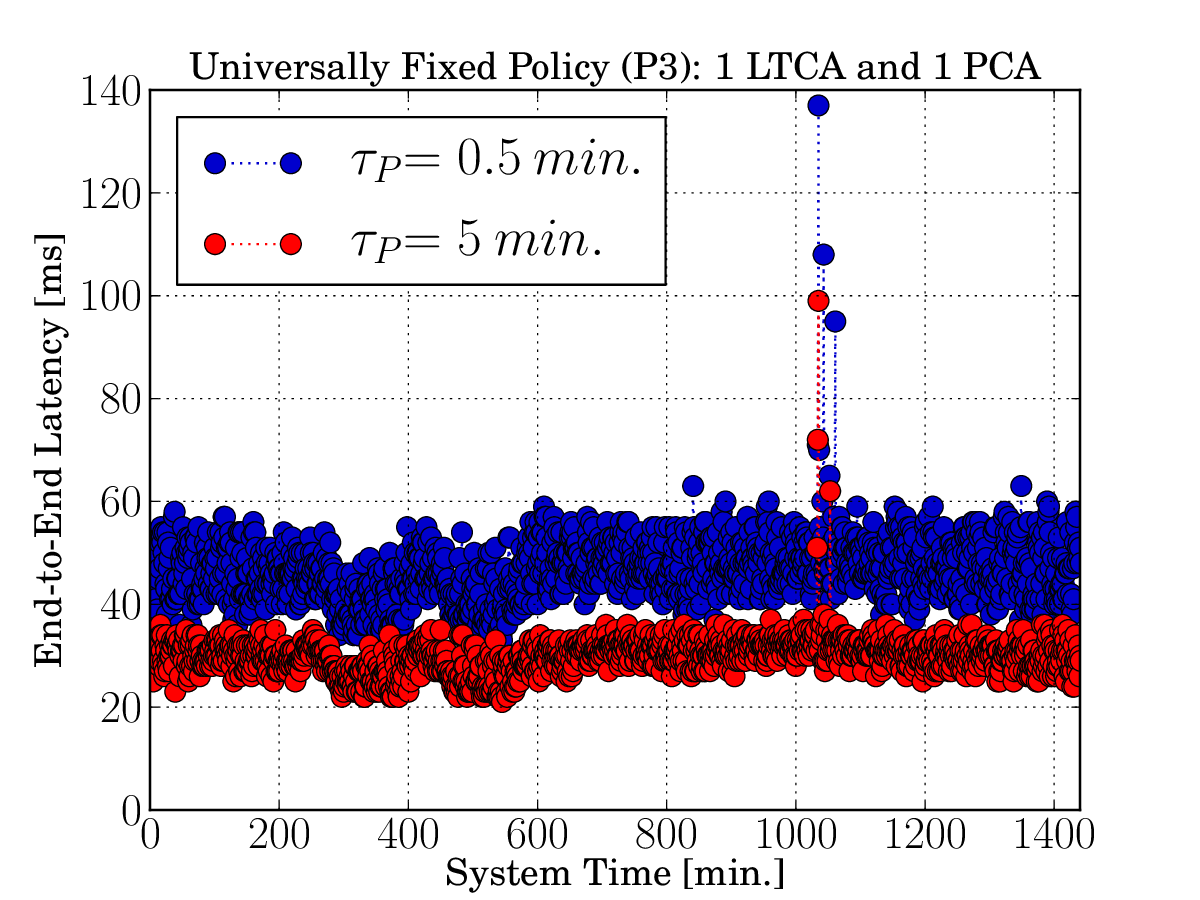}}
   \caption{End-to-end latency for P3, $\Gamma_{P3}$=5 minutes}
   \label{fig:mobihoc-2016-tapas_lust_entire_vehicle_delay_over_time_fixed_policy_optimized}
   \vspace{-1em}
\end{figure}

Figs. \ref{fig:mobihoc-2016-tapas_lust_entire_vehicle_delay_over_time_usercontrolled_policy}-\ref{fig:mobihoc-2016-tapas_lust_entire_vehicle_delay_over_time_fixed_policy_optimized} show the interplay between the end-to-end latency, averaged over all completed protocol executions within each minute period, and different pseudonym acquisition policies (with different configurations) for the two datasets. Table \ref{table:mobihoc-2016-policies-delays-information} details end-to-end latency statistics for each policy. With P1 (Fig. \ref{fig:mobihoc-2016-tapas_lust_entire_vehicle_delay_over_time_usercontrolled_policy}), each vehicle requests all required pseudonyms at once; with $\tau_{P}=0.5$ min., 99\% of the requesters for TAPASCologne and \ac{LuST} datasets are served within less than 153 ms and 167 ms respectively. As it is shown in Fig. \ref{fig:mobihoc-2016-tapas_lust_entire_vehicle_delay_over_time_usercontrolled_policy}.b, the end-to-end latency with P1 follows the arrival distribution and it is fluctuating over time; the reason is that with P1, vehicles can request for any trip duration, thus requesting more pseudonyms at once. 

With P2 (Fig. \ref{fig:mobihoc-2016-tapas_entire_vehicle_delay_over_time_upperlimit_policy}), the vehicles request a fixed amount of pseudonyms every time (for a duration of $\Gamma_{P2}$=5 min.), thus never overloading the \ac{PCA} server with large amount of pseudonyms acquisition in a single request; this results in a low standard deviation and variance, and a smooth end-to-end latency in compare with other policies. The end-to-end latency for the TAPAS and \ac{LuST} datasets ($\tau_{P}$=0.5 min.) is 50 ms and 45 ms respectively; accordingly, 99\% of vehicles are served within less than 109 ms and 80 ms respectively. 

With P3 (Fig. \ref{fig:mobihoc-2016-tapas_lust_entire_vehicle_delay_over_time_fixed_policy_optimized}), the system enforces synchronized batch arrivals to obtain pseudonyms: each vehicle requests pseudonyms for the entire $\Gamma_{P3}$, timely aligned with the rest. The end-to-end latency for the two datasets ($\tau_{P}$=0.5 min.) is 45 ms and 47 ms respectively; moreover, 99\% of the requesters are served within less than 70 ms and 74 ms respectively.

%#######################################
\begin{table}[!t]
	\caption{Latency Statistics for each Policy \protect\linebreak ($\Gamma_{}$ = 5 min., $\tau_{P}$ = 0.5 min.)}
	\resizebox{0.45\textwidth}{!}
		{
		\begin{minipage}{0.8\textwidth}
			\begin{tabular}{| l | |*{3}{c} |*{3}{c} |}
				\hline
				& \textbf{TAPAS-P1} & \textbf{TAPAS-P2} & \textbf{TAPAS-P3} & \textbf{\acs{LuST}-P1} & \textbf{\acs{LuST}-P2} & \textbf{\acs{LuST}-P3} \\\hline\hline
				\textbf{Maximum (ms)}   & 426 & 268 & 4254 & 504 & 248 & 3408 \\\hline 
				\textbf{Minimum (ms)}   & 17 & 26 & 18 & 15 & 25 & 20 \\\hline 
				\textbf{Average (ms)}   & 69 & 50 & 45 & 69 & 45 & 47 \\\hline 
				\textbf{Std. Deviation} & 26 & 17 & 23 & 30 & 12 & 21 \\\hline 
				\textbf{Variance}       & 708 & 295 & 535 & 895 & 138 & 449 \\\hline %($\mathbf{ms^2}$)
				$\mathbf{Pr\{t\leq x\}=0.99}$ \textbf{(ms)} & 153 & 109 & 70 & 167 & 80 & 74 \\\hline
			\end{tabular}
			\label{table:mobihoc-2016-policies-delays-information}
		\end{minipage} 
		}
\end{table}
%#######################################

%#######################################
%#######################################

%#######################################
%#######################################

%#######################################
%#######################################

The results confirm that our secure and privacy preserving scheme efficiently issues pseudonyms for the requesters; thus, an \ac{OBU} can initiate a request for pseudonyms within the lifetime of the last single valid pseudonym. We can conclude that modest \acp{VM} can serve very large number of vehicles even during the most harsh traffic conditions with very low delays, and the most promising policy in terms of privacy protection incurs moderate overhead. The choice of parameters for $\Gamma_{P2/P3}$ and $\tau_{P}$ mainly determines the frequency of interaction with the \ac{VPKI} and the volume of workload imposed to the \ac{PCA}: the shorter the pseudonym lifetimes are, the greater number of pseudonyms will be requested, thus a higher workload is imposed on the \ac{PCA}. As the results show, issuing pseudonyms with very short lifetimes (30 sec.) does not have a high impact on the overall performance of the servers. The results presented here are obviously dataset-dependent; however, by understanding the characteristics of the mobility, i.e., the road-constrained movements, the appearance of the \acp{RSU}, vehicle movement direction, and sudden bursts of traffic, system designers can evaluate the impact of a given mobility trace on the deployment and dimensioning of the \ac{VPKI} resources. 

\section{Discussion and Conclusion} 
\label{sec:mobihoc-2016-conclusions}

\textbf{Remark on the suitability of non-overlapping pseudonym lifetimes for safety applications:} As mentioned earlier, safety applications can operate more easily if there is linkability (with the vehicle keeping the same pseudonym) during a critical situation. In such a case, e.g., emergency braking or collision avoidance, the vehicle can simply include a link to its previous pseudonym. In fact, the vehicle could even sign with two $k^{i-1}_v$ and $k^{i}_v$ private keys, corresponding to  pseudonyms $P^{i-1}_v$ and $P^{i}_v$. This would ensure the operation of the safety application (partial linkability). 

\textbf{Summary and future work:} In this paper, we specified three policies for pseudonym acquisition, drawing from the literature. We integrated those into the pseudonym acquisition process of the state-of-the-art \ac{VPKI} system \cite{khodaei2014ScalableRobustVPKI}. To the best of our knowledge, our system \cite{khodaei2014ScalableRobustVPKI} is the latest and fastest \ac{VPKI}. Nonetheless, our investigation is relevant to any \ac{VPKI} that relies on non-overlapping on-demand pseudonyms acquisition. We presented a secure and efficient solution for pseudonyms acquisition while the timing information cannot harm user privacy. Through experimental evaluation, we demonstrated that modest \acp{VM} dedicated as servers can serve on-demand requests with very low delay, and the most promising policy in terms of privacy protection incurs moderate overhead. 

Using P1, a vehicle interacts with the \ac{VPKI} servers once to obtain the necessary pseudonyms for the entire trip duration (ideally without over-provisioning). However, according to P2 and P3, vehicles could be potentially equipped with more pseudonyms than needed, i.e., the \ac{PCA} might issue pseudonyms for a period during which the vehicle will not use them. In general, the longer the pseudonym refill interval ($\Gamma_{P2}$ or $\Gamma_{P3}$) is, the less frequent vehicles-\ac{VPKI} interactions, but the higher the chance to overprovision a vehicle. As future work, we will investigate the pseudonym utilization with various configurations ($\Gamma_{P2/P3}$ and $\tau_{P}$) to investigate the interplay with the server workload and privacy protection (the shorter $\tau_{P}$, the less linkable are messages by a vehicle). We further intend to rigorously analyze the security and privacy protocols and evaluate the level of privacy, i.e., unlinkability, based on the timing information of the pseudonyms for each policy.

\bibliographystyle{abbrv}
\bibliography{references}  % sigproc.bib is the name of the Bibliography in this case

% that's all folks
\end{document}